# Novel polymorphic phase of BaCu$_2$As$_2$: impact of flux for new phase formation in crystal growth

Hanlin Wu,[†] Sheng Li,[†] Zheng Wu,[‡] Xiqu Wang,[§] Gareth A. Ofenstein,[†] Sunah Kwon,[||] Moon J. Kim,[||] Paul C. W. Chu*,[‡] and Bing Lv*,[†]

[†] Department of Physics, The University of Texas at Dallas, Richardson, TX 75080, USA
[‡] Texas Center for Superconductivity (TcSUH) and Department of Physics, University of Houston, Houston, Texas, 77204, USA
[§] Department of Chemistry, University of Houston, Houston, TX 77204, USA
[||] Department of Materials Science and Engineering, University of Texas at Dallas, Richardson, TX 75080, USA

**ABSTRACT**
In this work, we have thoroughly studied the effects of flux composition and temperature on the crystal growth of the BaCu$_2$As$_2$ compound. While Pb and CuAs self-flux produce the well-known α-phase ThCr$_2$Si$_2$-type structure (Z=2), a new polymorphic phase of BaCu$_2$As$_2$ (β phase) with a much larger *c* lattice parameter (Z=10), which could be considered an intergrowth of the ThCr$_2$Si$_2$- and CaBe$_2$Ge$_2$-type structures, has been discovered *via* Sn flux growth. We have characterized this structure through single-crystal X-ray diffraction, transmission electron microscopy (TEM), and scanning transmission electron microscopy (STEM) studies. Furthermore, we compare this new polymorphic intergrowth structure with the α-phase BaCu$_2$As$_2$ (ThCr$_2$Si$_2$ type with Z=2) and the β-phase BaCu$_2$Sb$_2$ (intergrowth of ThCr$_2$Si$_2$ and CaBe$_2$Ge$_2$ types with Z=6), both with the same space group *I4/mmm*. Electrical transport studies reveal *p*-type carriers and magnetoresistivity up to 22% at 5 K and under a magnetic field of 7 T. Our work suggests a new route for the discovery of new polymorphic structures through flux and temperature control during material synthesis.

**INTRODUCTION**

The discovery of Fe-based superconductors[1] has spurred worldwide efforts not only to understand the interplay between magnetism and superconductivity but also to search for new types of superconductors with similar structures. Among them, the so-called 122 family, in which the parent compounds have the composition AFe$_2$As$_2$ (A = Ca, Sr, Ba, Eu, K, Rb, and Cs) with the tetragonal ThCr$_2$Si$_2$-type structure and space group *I4/mmm*, is probably the most studied and exploited[2-9] due to the availability of high-quality single crystals allowing a broad range of advanced characterization and the large materials base in this particular crystal-structure family. Many techniques have been developed to grow the crystals, and among them, high-temperature flux growth is widely used for obtaining single crystals of 122-family compounds. Different types of flux, from metal flux, such as Sn[10] and In[11,12], to self-flux (*i.e.*, the flux is consumed and is partially incorporated into the resulting crystals), such as FeAs[13] and KAs[14], are used. However, single crystals grown by metal flux sometimes show different properties than those obtained from self-flux. For instance, the spin density wave transition is significantly suppressed in the Sn-flux-grown BaFe$_2$As$_2$[10,15] due to the inclusion of small amounts of Sn into the lattice. BaFe$_2$As$_2$ crystals grown from In flux[11,12] even show a superconducting transition with a full drop in resistivity but without diamagnetic screening at various values of T$_c$ (19-23 K). Whether such superconductivity is induced by the incorporation of In into the lattice remains unclear at this moment. Annealing studies indicate that such a transition is not related to lattice distortion or strain, in contrast to similar superconducting behavior observed in FeAs-flux-grown SrFe$_2$As$_2$[16]. X-ray crystallography, when paired with precise chemical analysis, is an effective tool[15,17-19] for unambiguously determining the quantity and location of Sn in the lattice.

In order to obtain further insight into the physics of Fe pnictides, much research effort has also been devoted to searching for superconductivity in the Fe-free pnictide systems with similar structures, initially in those with ThCr$_2$Si$_2$-type structures and later expanding to those with the variant CaBe$_2$Ge$_2$- and BaAl$_4$-type structures.[20-37] In fact, many new superconductors have been discovered in these families, but with a different Fermi surface topology than the Fe-based superconductors.[20,21,23,27-31] The Cu-based 122 pnictides are another family of compounds that initially attracted attention since they could possibly serve as a bridge between the cuprate and Fe-based superconductors.[38-42] Unfortunately, no superconductivity has been discovered in these compounds thus far. Both calculations and angle-resolved photoemission spectroscopy studies of BaCu$_2$As$_2$ have shown that its Cu 3d bands are far below the Fermi level, and thus have a weak electronic correlation.[43-46] Nevertheless, this family of compounds is rather interesting from a structural perspective. The As-based compounds ACu$_2$As$_2$ (A = Ca, Sr, Ba, and Eu) all crystallize with a body-centered tetragonal ThCr$_2$Si$_2$ structure, but the Sb-based compounds ACu$_2$Sb$_2$ (A = Sr, Ba, and Eu) mainly crystallize in the related primitive-tetragonal CaBe$_2$Ge$_2$-type structure.[47-49] For BaCu$_2$Sb$_2$, a very interesting intergrowth structure with ordered ThCr$_2$Si$_2$ and CaBe$_2$Ge$_2$ types of unit cells has been reported.[38,39] The formation of such an intergrowth structure appears to be independent of flux (self-flux or Pb flux), but rather depends on the synthetic temperature profile. This finding triggered our interest in exploring the effects of flux and temperature on BaCu$_2$As$_2$ crystal growth. Here we report on a new polymorphic phase of BaCu$_2$As$_2$ that is different from the BaCu$_2$Sb$_2$ polymorph and has a much larger $c$ lattice parameter (51.05 Å compared to 32.6 Å). This new structure has been unambiguously verified by single-crystal diffraction, TEM, and STEM studies. The effects of starting materials, flux, temperature profiles, and possible flux inclusion into the lattice were also thoroughly studied. Our results suggest that the discovery of new polymorphic phases is possible through carefully controlled flux conditions. Electrical transport studies of this new phase reveal that it exhibits $p$-type carriers and magnetoresistivity up to 22% at 5 K and under a magnetic field of 7 T.

**EXPERIMENTAL SECTION**

**Materials synthesis and single-crystal growth**

Ba rods (99.5%), Cu powder (99.99%), As lumps (99.999%), Sn shots (99.99%) and Pb ingots (99.999%) from Alfa Aesar were used as received for our reactions. Three different fluxes were used for our crystal growth: CuAs, Sn, and Pb. The X-ray powder pure CuAs precursor was synthesized through heating a stoichiometric mixture of Cu and As in a sealed quartz tube at 800 °C for two days, and in two iterations. The ThCr$_2$Si$_2$-type α-phase BaCu$_2$As$_2$ single crystals were grown by either the self-flux method using CuAs as the flux with Ba:CuAs in a molar ratio of 1:4 or the Pb flux method with molar ratio Ba:Cu:As:Pb = 1:2:2:10. In each case, the materials were sealed in an evacuated quartz tube using an alumina crucible as a container and subsequently heated up overnight to a target temperature. For CuAs flux, the assembly was heated up to 1080 °C for 24 h, and then slowly cooled at a rate of 3 °C/h down to 800 °C. For the Pb flux, the assembly was heated up to 1100 °C for 24 h and then slowly cooled at a rate of 3 °C/h down to 600 °C. The new polymorphic β phase of BaCu$_2$As$_2$ was synthesized using Sn flux, in which Ba:CuAs:Sn in a 1:4:30 molar ratio was sealed in an evacuated quartz tube with an alumina crucible, heated up to 1100 °C for 24 h, and then slowly cooled at a rate of 3 °C/h down to 550 °C. Large plate-like crystals with a typical size of 3 × 3× 0.5 mm$^3$ were obtained by decanting the flux with a centrifuge at high temperature (800 °C for CuAs flux, 600 °C for Pb flux, and 550 °C or 500 °C for Sn flux) using quartz wool as a filter. It should be noted that needle-like BaCu$_{10}$As$_4$ crystals are also produced as byproducts during CuAs-flux growth and can be easily separated from the plate-like α-phase BaCu$_2$As$_2$ due to their different morphologies. To fully investigate the growth conditions for the new β-phase BaCu$_2$As$_2$, several growth parameters (*e.g.* precursor materials and their concentrations) and growth temperature profiles have been tested, with results provided below.

**Characterization**

X-Ray Diffraction: Powder X-ray diffraction (XRD) was performed using a Rigaku Smartlab diffractometer with Cu K$_\alpha$ radiation. Single-crystal X-ray Diffraction was performed using a Bruker SMART diffractometer equipped with an Apex II area detector and an Oxford Cryosystems 700 Series temperature controller at room temperature. A hemisphere of frames was measured using a narrow-frame method with a scan width of 0.30° in ω and an exposure time of 60 s/frame with Mo K$_\alpha$ radiation. The collected dataset was integrated using the Bruker Apex-II program, with the intensities corrected for the Lorentz factor, polarization, air absorption, and absorption due to variation in the path length through the detector faceplate. The data were scaled, and absorption correction was applied using SADABS.[50] A starting model was obtained using the intrinsic method in SHELXT[51], and atomic sites were refined anisotropically using SHELXL2014.

Chemical Analysis: Chemical analyses were performed using both energy-dispersive X-ray spectroscopy (EDS) and wavelength-dispersive spectrometry (WDS) on a JEOL JXA-8600 electron microprobe analyzer. Data were collected at multiple (> 5) points for each sample to ensure accuracy. The instrumental error of these results is smaller than 0.5%.

TEM and STEM: Atomic-resolution high-angle annular dark-field scanning transmission electron microscopy (HAADF-STEM) images, high-resolution TEM (HRTEM), and selected area electron diffraction (SAED) were performed using an aberration-corrected JEM-ARM200F (JEOL USA Inc.) operated at 200 kV. SAED simulation was performed by JEMS software. TEM lamellae were prepared using a FEI Nova 200 dual-beam focused-ion-beam system (FIB, FEI Inc.) by the lift-out method.

Electrical Transport and Magnetic Studies: Temperature (T)- and magnetic field (H)-dependent electrical resistivity ρ(T, H) using a four-probe configuration, and Hall resistivity with a five-probe configuration, were measured down to 5 K and up to 7T magnetic field in a Quantum Design Physical Property Measurement System (PPMS). Gold wires, each with a radius of 20 μm, were used as the electrical leads for the measurements and were attached to the crystal surfaces using silver paste with contact resistance normally of ~ 1-5 Ω. Temperature dependent magnetic susceptibility measurement were carried out down to 5K at 5T magnetic field in a Quantum Design Magnetic Property Measurement System (MPMS).

**RESULTS AND DISCUSSION**

The new β phase of BaCu$_2$As$_2$ was initially identified from XRD scanning of the plate-shaped crystals grown by three different fluxes. The XRD diffraction patterns are highly *c*-axis-oriented due to the preferred orientation of the crushed single crystals, as expected. We observed a few additional small peaks in the XRD patterns of the Sn-flux-grown crystals and initially attributed them to misalignment of the crystals or possible impurities. However, chemical analysis from both EDS and WDS measurements clearly indicated the existence of only elemental Ba, Cu, and

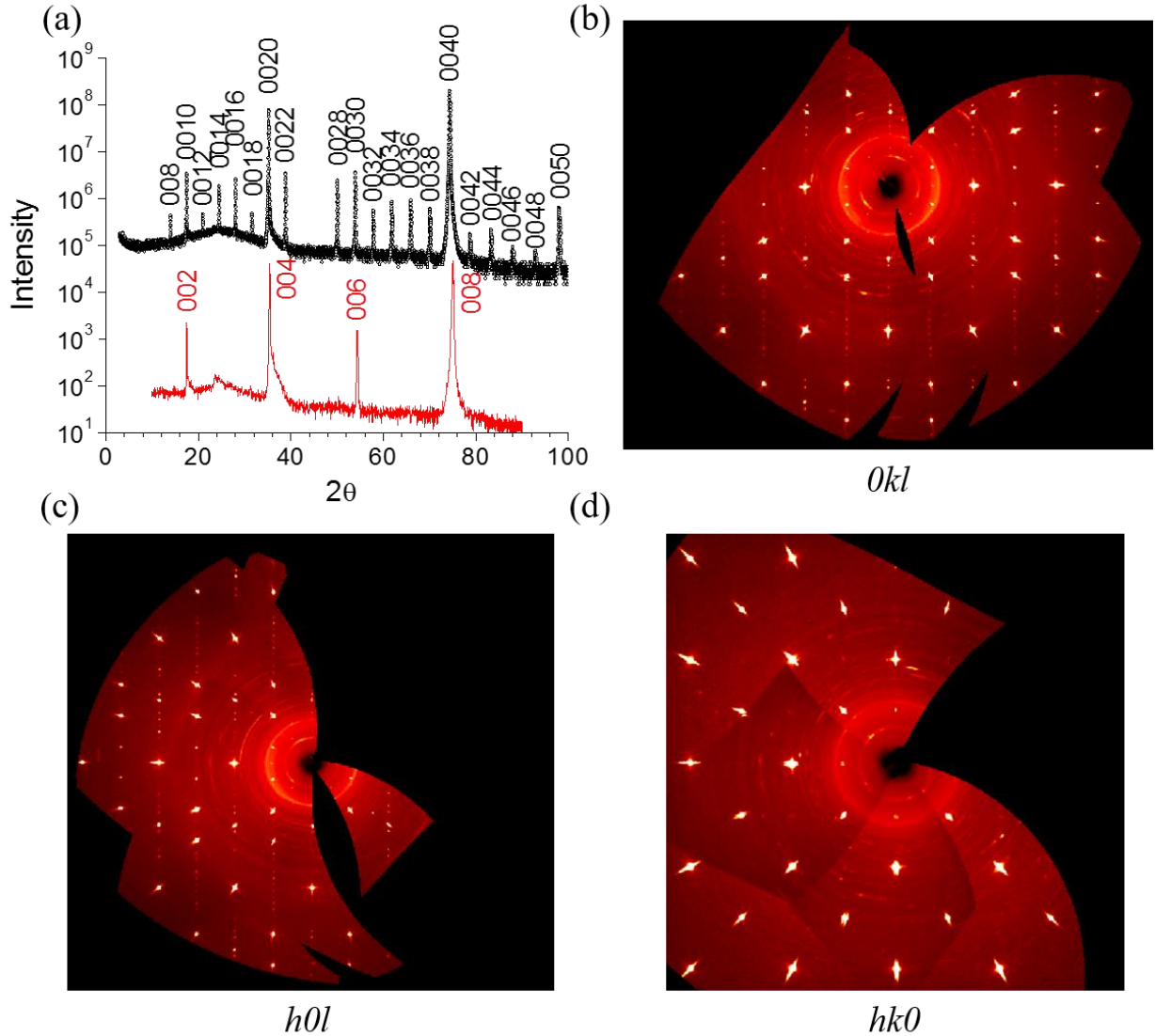

Fig. 1. (a) XRD patterns of α-BaCu$_2$As$_2$ (red) and β-BaCu$_2$As$_2$ (black) along the (*00l*) orientation using the logarithmic scale; the intensity of the β-BaCu$_2$As$_2$ pattern is offset along the ordinate axis for better comparison. (b), (c) and (d): Integrated precession images from the full X-ray single-crystal data collection of β-BaCu$_2$As$_2$ along (*0kl*), (*h0l*), and (*hk0*), respectively, showing the superstructure along the *c* axis.

As in the crystals, and neither impurities nor flux inclusions were detected within the instrument detection limits. Furthermore, the (*00l*) Miller indices of these additional peaks suggest an ordered crystal lattice, implying the presence of a superstructure in these Sn-flux-grown crystals, as shown in the logarithmic scale in Fig. 1a (black pattern). For the CuAs- and Pb-flux-grown crystals, we did not observe these peaks (red pattern in Fig. 1a), suggesting that these crystals indeed adopted a conventional ThCr$_2$Si$_2$-type structure. We therefore employed X-ray single-crystal diffraction to investigate the superstructure and fully understand the structure of this new phase in detail. A fast omega scan using a narrow-frame algorithm with an exposure time of 30 seconds/frame initially leads to the assignment of the ThCr$_2$Si$_2$-type unit cell. However, under a full set of data collection with a longer exposure time of 50 seconds/frame, all Sn-flux-grown crystals clearly show a superstructure with a much larger $c$ axis of 51.05(1) Å, while the CuAs- and Pb-flux-grown crystals have the ThCr$_2$Si$_2$-type structure with $c$= 10.073(6) Å. Crystallographic parameters and refinement details are provided in Table 1. Atomic coordinates, anisotropic displacement parameters, occupancies, and some selected interatomic distances and angles are provided in Table 2. The integrated precession images of the (*0kl*), (*h0l*), and (*hk0*) layers from the complete set of single-crystal diffraction data, as shown in Fig. 1b-d, provide clear visualization of the existence of the superstructure in our new phase.

Table 1. Crystallographic data for α-BaCu$_2$As$_2$ and β-BaCu$_2$As$_2$.

| Empirical formula | α-BaCu$_2$As$_2$ (ThCr$_2$Si$_2$ type) | β-BaCu$_2$As$_2$ |
|---|---|---|
| Space group | *I4/mmm* (No.139) | *I4/mmm* (No.139) |
| Unit cell dimensions | a= 4.445(2) Å, c=10.073(6) Å | a= 4.425(1) Å, c=51.05(1) Å |
| Unit cell volume | 199.0(2) Å$^3$ | 999.7(4) Å$^3$ |
| Z | 2 | 10 |
| Absorption coefficient | 36.629 mm$^{-1}$ | 36.462 mm$^{-1}$ |
| F(000) | 360 | 1800 |
| Crystal size | 0.11 x 0.08 x 0.05 mm$^3$ | 0.08 x 0.08 x 0.01 mm$^3$ |
| Theta range for data | 4.05° to 29.18° | 1.596° to 28.288° |
| Reflections collected | 494 | 3140 |
| Independent reflections | 100[$R_{int}$ = 0.0342] | 445[$R_{int}$ = 0.0327] |
| Max. and min. transmission | 0.746, 0.355 | 0.746, 0.287 |
| Data / restraints / parameters | 100 / 0 / 9 | 445 / 0 / 35 |
| Goodness-of-fit on F$^2$ | 1.225 | 1.136 |
| Final R indices [I > 2σ] | $R_1$ = 0.0308, $wR_2$ = 0.0813 | $R_1$ = 0.0378, $wR_2$ = 0.1250 |
| R indices (all data) | $R_1$ = 0.0311, $wR_2$ = 0.0815 | $R_1$ = 0.0365, $wR_2$ = 0.1235 |
| Extinction coefficient | 0.016(4) | 0.0014(2) |
| Largest diff. peak and hole | 1.68/-1.26 | 2.04/ -1.53 |

Table 2. Atomic coordinates, selected interatomic distances, and equivalent isotropic displacement parameters ($U_{eq}$) of α-BaCu$_2$As$_2$ and β-BaCu$_2$As$_2$. $U_{eq}$ is defined as 1/3 of the trace of the orthogonalized $U_{IJ}$ tensor.

| Type | Atom | Wyckoff site | *Symm.* | x/a | y/b | z/c | Occ. | $U_{eq}$ (Å$^2$) |
|---|---|---|---|---|---|---|---|---|
| α | As | 4e | *4mm* | 1/2 | 1/2 | 0.1277(1) | 1 | 0.0105(5) |
| | Cu | 4d | *-4m2* | 0 | 1/2 | 1/4 | 1 | 0.0177(8) |
| | Ba | 2a | *4/mmm* | 0 | 0 | 0 | 1 | 0.0143(5) |
| β | As1 | 4e | *4mm* | 1/2 | 1/2 | 0.02537(2) | 1 | 0.0111(4) |
| | As2 | 4e | *4mm* | 0 | 0 | 0.07412(2) | 1 | 0.0107(4) |
| | As3 | 4e | *4mm* | 0 | 0 | 0.12441(2) | 1 | 0.0094(4) |
| | As4 | 4e | *4mm* | 1/2 | 1/2 | 0.17481(2) | 1 | 0.0086(4) |
| | As5 | 4d | *-4m2* | 1/2 | 1 | 1/4 | 1 | 0.0089(4) |
| | Ba1 | 2a | *4/mmm* | 0 | 0 | 0 | 1 | 0.0111(4) |
| | Ba2 | 4e | *4mm* | 1/2 | 1/2 | 0.10086(2) | 1 | 0.0097(4) |
| Ba3 | 4e | *4mm* | | 0 | 0 | 0.19948(2) | 1 | 0.0088(4) |
| Cu1 | 8g | *2mm.* | | 0 | 1/2 | 0.05008(2) | 1 | 0.0212(4) |
| Cu2 | 8g | *2mm.* | | 0 | 1/2 | 0.14997(2) | 1 | 0.0166(4) |
| Cu3 | 4e | *4mm* | | 1/2 | 1/2 | 0.22486(2) | 1 | 0.0134(4) |
| Selected interatomic distances (Å) | α | Cu1-As1 | 2.541(1) | | | | As1-As1 | 2.572(3) |
| | β | Cu1-As1 | 2.547(1) | Cu2-As4 | 2.550(1) | | As1-As1 | 2.590(2) |
| | | Cu1-As2 | 2.530(1) | Cu3-As5 | 2.557(1) | | As2-As3 | 2.567(1) |
| | | Cu2-As3 | 2.569(1) | | | | | |

The refined crystal structure of the new β-phase $BaCu_2As_2$ is shown in Fig. 2. For purposes of comparison, we also plotted the α phase of $BaCu_2As_2$ with the $ThCr_2Si_2$-type structure (Fig. 2a) and a hypothetical $BaCu_2As_2$ with the closely related $CaBe_2Ge_2$-type structure (Fig. 2b). Both structures contain two tetragonally coordinated $Cu_2As_2$ layers per unit cell sandwiched by the Ba atoms. In the $ThCr_2Si_2$-type structure, the neighboring $Cu_2As_2$ layers are equivalent and are related by inversion symmetry, which results in a body-centered lattice. The As-As distances between the two identical $Cu_2As_2$ layers are relatively short, suggesting significantly strong interactions between the $Cu_2As_2$ layers, akin to the collapsed tetragonal phases discovered in the Fe-pnictide superconductors[52-55]. $CaBe_2Ge_2$-type structures are instead made up of alternating $Cu_2As_2$ and $As_2Cu_2$ layers, and thus break both the mirror and inversion symmetries, resulting in a primitive lattice. The new β phase of $BaCu_2As_2$ could be considered an ordered intergrowth of $ThCr_2Si_2$-type (block A motif) and $CaBe_2Ge_2$-type (block B motif) structures. The unit cell of our β-phase $BaCu_2As_2$ consists of five of these unit-cell blocks stacked along the *c* axis, where $CaBe_2Ge_2$-type blocks are sandwiched between $ThCr_2Si_2$-type blocks with an ABABA stacking pattern (Fig. 2c). Interestingly, this stacking pattern retains the mirror symmetry, resulting in an overall space group symmetry of *I4/mmm*, the same as for the $ThCr_2Si_2$-type structure. The position exchange of Cu (Wyckoff 4d site) and As (Wyckoff 4e site), which swap positions from the α-phase $ThCr_2Si_2$-type structure to the $CaBe_2Ge_2$-type structure, and the overall ABABA stacking lead to eleven unique crystallographic sites in our β-phase $BaCu_2As_2$, in comparison to only three unique positions in the α-phase $BaCu_2As_2$. We have paid close attention to the possibility of Sn-flux inclusion in the crystal lattice during our single-crystal studies, as has been seen in iron-based superconductors grown from Sn flux[10]. The full occupancy at each crystallographic site in our crystal refinement excludes such a possibility. This observation, together with the chemical analysis results from EDS and WDS measurements, leads us to conclude that flux inclusion during the flux growth of Cu-based 122 compounds is unlikely, which is consistent with other reports.

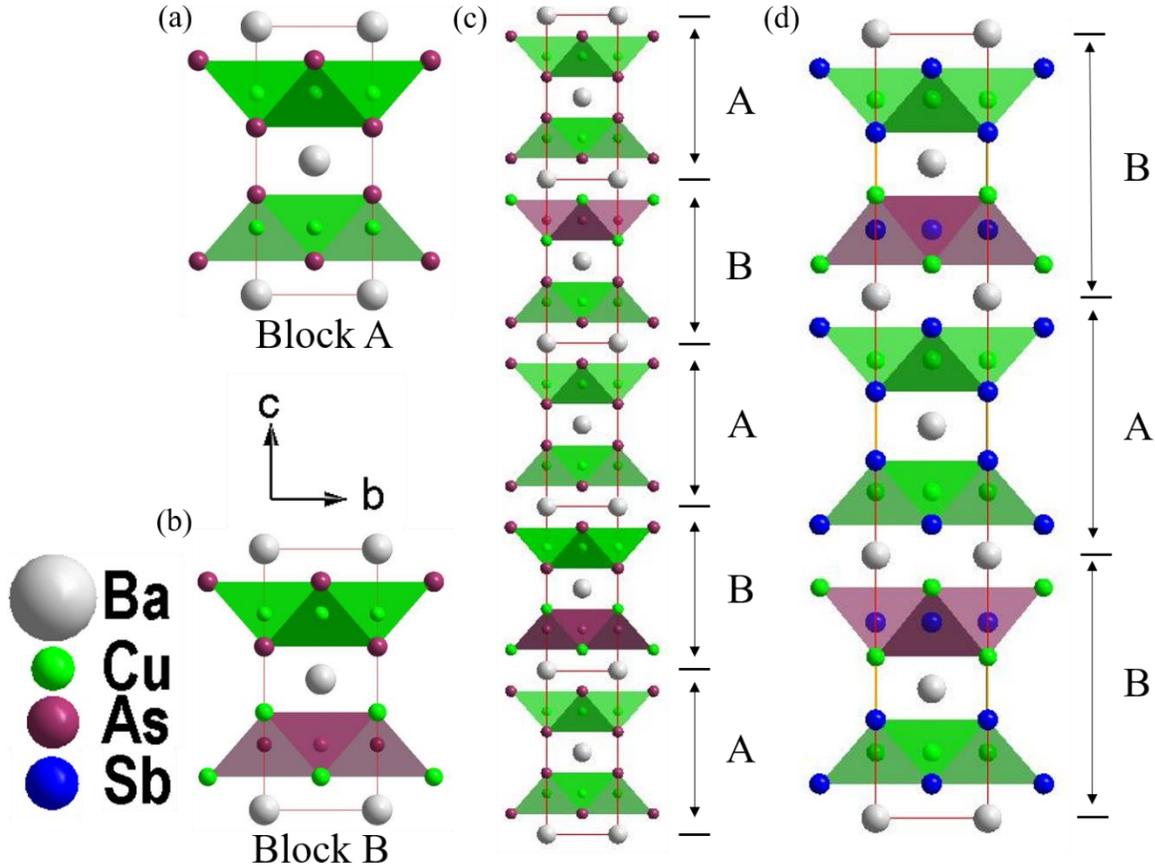

Fig. 2. (a) α-BaCu$_2$As$_2$ with the ThCr$_2$Si$_2$-type (block A motif) structure; (b) α-BaCu$_2$As$_2$ with a hypothetical CaBe$_2$Ge$_2$-type (block B motif) structure; (c) intergrowth structure of β-BaCu$_2$As$_2$ with the ABABA stacking pattern; and (d) intergrowth structure of β-BaCu$_2$Sb$_2$ with the BAB stacking pattern.

It is also worth mentioning that a different intergrowth stacking pattern of ThCr$_2$Si$_2$-type (block A motif) and CaBe$_2$Ge$_2$-type (block B motif) structures has also been reported in the Sb-analog β-BaCu$_2$Sb$_2$.[39,40] Different from our β-BaCu$_2$As$_2$, the β-BaCu$_2$Sb$_2$ consists of three unit-cell blocks with the same overall symmetry $I4/mmm$, where a ThCr$_2$Si$_2$-type block is sandwiched between two CaBe$_2$Ge$_2$-type blocks with a BAB stacking pattern, as illustrated in Fig. 2d.

To further elaborate the structural difference between the α and β phases of BaCu$_2$As$_2$ crystals, HRTEM and HAADF-STEM images were obtained. SAED and STEM imaging were carried out along the [100] zone axis. All SAED simulations were conducted using constructed models that were built based on the single-crystal diffraction results. Acquired TEM and SAED patterns are both consistent with the simulated patterns, indicating that the constructed models accurately represent the single crystals. Fig. 3a, 3b show the acquired and simulated SAED patterns, respectively of α-BaCu$_2$As$_2$, and Fig. 3d, 3e show these respective SAED patterns for β-BaCu$_2$As$_2$. It can be clearly seen in the simulated SAED pattern of β-BaCu$_2$As$_2$ shown in Fig. 3d that there are not only {001} fundamental spots but also additional 1/5{00$l$} superstructure reflections, which are not observed for α-BaCu$_2$As$_2$ (Fig. 3b), further supporting the single-crystal diffraction results. Fig. 3c shows the HAADF-STEM image taken from an α-BaCu$_2$As$_2$ single crystal along the [001] zone axis. The STEM images were filtered to enhance contrast by removing noise. The existence of each element was confirmed by STEM and EDX analyses, from which we also observed the lack of flux inclusions in our crystals. In Z-contrast HAADF images, each atom can be distinguished easily by brightness and size. Ba atoms are clearly recognized as bright dots, and Cu atoms are dimmer than As atoms. For comparison, STEM images from a β-$BaCu_2As_2$ single crystal were obtained and are shown in Fig 3f. Each constructed model is overlapped with the STEM images to confirm their consistency. Additional As planes are clearly observed in the $BaCu_2As_2$ superstructure and are marked yellow in Fig. 3f, which also shows the intergrowth superstructure in β-$BaCu_2As_2$ consisting of two $CaBe_2Ge_2$-type unit cell blocks and three $ThCr_2Si_2$-type unit cell blocks, which is consistent with our X-ray single-crystal diffraction and SAED pattern results.

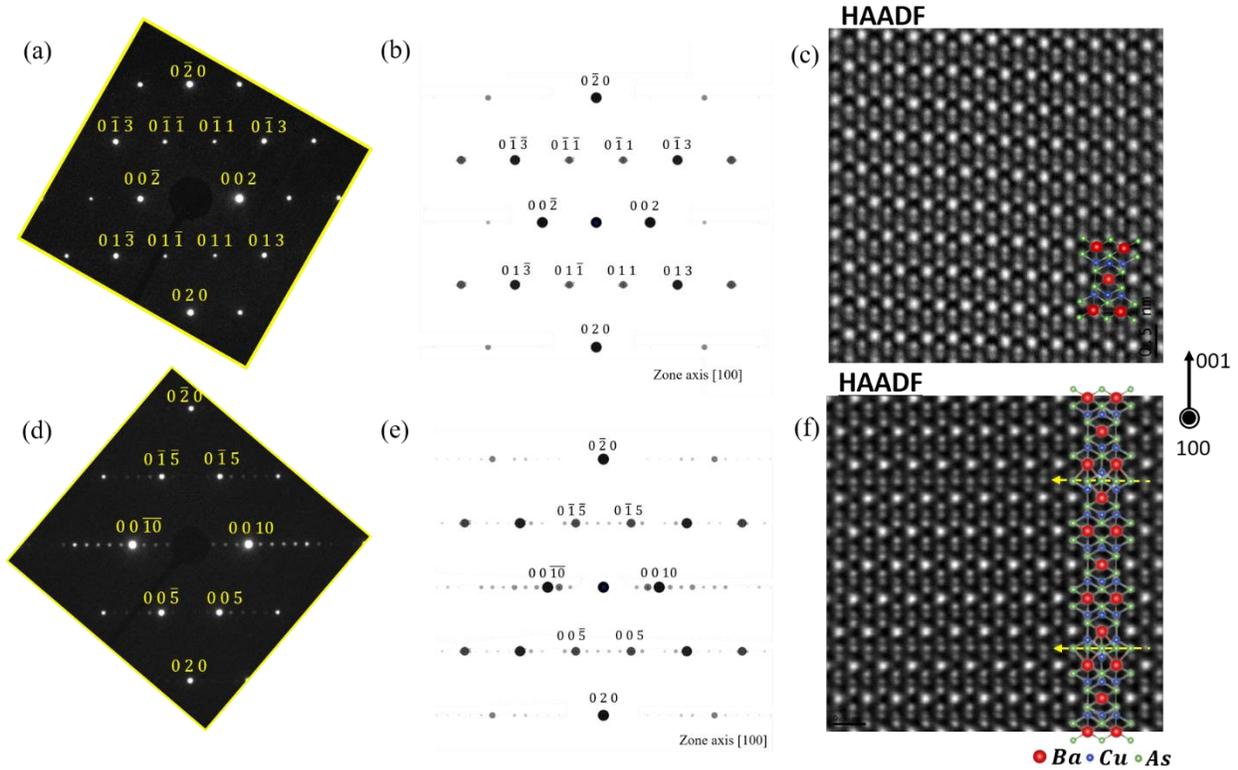

Fig. 3. (a) Acquired and (b) simulated SAED patterns and (c) STEM image of α-$BaCu_2As_2$. (d) Acquired and (e) simulated SAED patterns and (f) STEM image of β-$BaCu_2As_2$.

To fully understand the effect of flux on the formation of different polymorphic phases, to determine the growth window for this new β-$BaCu_2As_2$ phase, and to further explore possible new phases using flux growth, we have systematically studied the effects of the precursors, temperature, and materials/flux ratios in a series of controlled experiments with some of the key results summarized in Table 3. We found the following: 1) Using CuAs precursor is preferred to elemental Cu and As for the crystal growth when using Sn flux. Regardless of the starting ratios, using Ba:Cu:As:Sn generally produces the known air-sensitive ternary compound $Ba_3Sn_2As_4$[56] rather than the $BaCu_2As_2$ phase. 2) The starting Ba:CuAs ratio needs to be less than 1:2 to obtain $BaCu_2As_2$ crystals rather than $Ba_3Sn_2As_4$ crystals during Sn-flux growth. A Ba:CuAs ratio of 1:4 is optimal to consistently obtain high-quality $BaCu_2As_2$ phase. 3) β-$BaCu_2As_2$ phase formation is insensitive to the growth temperature (at least for the temperature range studied here). We found that growth temperatures ranging from 800 ºC to 1100 ºC for a Ba:CuAs:Sn ratio of 1:4:30 all result in the $\beta$-$BaCu_2As_2$ phase with the same superstructure features based on X-ray diffraction results. 4) The starting materials are not important for Pb-flux growth. Both elemental Cu/As and CuAs precursors are able to produce the α-phase $BaCu_2As_2$. On the other hand, attempts to use Ga as flux under similar synthesis conditions did not produce any crystals. Clearly, Sn flux has a compelling advantage over other fluxes for the growth of $\beta$-$BaCu_2As_2$, but the synthetic temperature does not make a difference in the phase formation once the flux is chosen. These results are different from the observation for $BaCu_2Sb_2$, where it is the temperature profile rather than the flux that plays an important role in determining α- or β-phase crystal growth. This further suggests that potential new phases, especially the intergrowth of $ThCr_2Si_2$-type and $CaBe_2Ge_2$-type structures that we observed here could be further explored in other 122 compounds through careful modification of both different types of flux (or their joint flux) and synthetic temperature profiles.

Table 3. Growth conditions for α-$BaCu_2As_2$ and β-$BaCu_2As_2$ single crystals.

| Element ratio [Ba:CuAs(or Cu:As): flux] | Flux | Temperature (°C) (Initial/Centrifuge) | Products |
|---|---|---|---|
| 1:1:10 | Sn | 950/500 | $Ba_3Sn_2As_4$[56] |
| 1:2:10 | Sn | 950/500 | $Ba_3Sn_2As_4$[56] |
| 1:2:2:10 (Cu:As) | Sn | 950/500 | $Ba_3Sn_2As_4$[56] |
| 1:2:30 | Sn | 950/550 | $Ba_3Sn_2As_4$[56] |
| 1:4:30 | Sn | 950/500 | $\beta$-$BaCu_2As_2$(Z=10) |
| 1:4:30 | Sn | 1100/550 | $\beta$-$BaCu_2As_2$(Z=10) |
| 1:2:10 (CuAs) | Pb | 1100/600 | $\alpha$-$BaCu_2As_2$(Z=2) |
| 1:2:2:10 (Cu:As) | Pb | 1100/600 | $\alpha$-$BaCu_2As_2$(Z=2) |
| 1:4 | CuAs | 1080/800 | $\alpha$-$BaCu_2As_2$(Z=2) |
| 1:2:30 | Ga | 900/500 | No crystal |

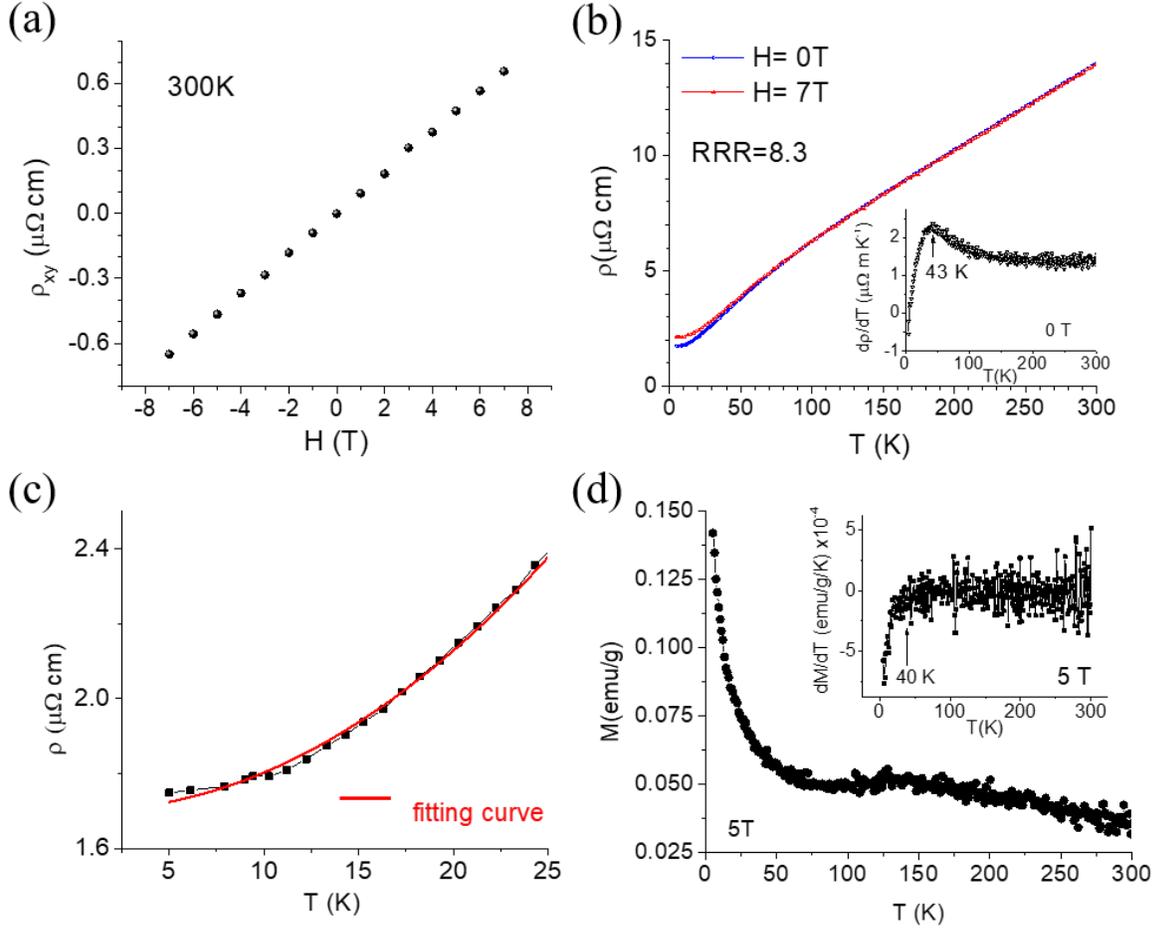

Fig. 4. (a) Hall resistivity data at room temperature for β-BaCu$_2$As$_2$. (b) Resistivity data of β-BaCu$_2$As$_2$ from 5 K to 300 K under magnetic fields of 0 T and 7 T. The inset shows first derivation of resistivity data under magnetic field of 0 T. (c) Resistivity data at low temperature range between 5 and 25K under a magnetic field of 0 T, together with a fitting curve (solid red line). (d) Magnetic susceptibility data of β-BaCu$_2$As$_2$ from 5K to 300K under magnetic field of 5 T. The inset is first derivation of susceptibility data, which shows an anomaly at 40 K.

Both field-dependent transverse Hall resistivity, temperature-dependent in-plane longitudinal resistivity measurement and magnetic susceptibility measurement were carried out on this new polymorphic β-BaCu$_2$As$_2$, as shown in Fig. 4. The room-temperature Hall resistivity suggests hole-like carriers in this system, and the estimated carrier concentration is about 6.63×10$^{21}$/cm$^3$ based on a single band model (Fig. 4a). The temperature-dependent in-plane resistivity data show overall metallic behavior, as expected, with a reasonably high residual resistivity ratio (RRR) ~8.3 compared to the other Cu-based 122 crystals[38,39] and a residual resistivity of 1.8 μΩ cm, relatively low among the 122-type compounds,[27-29, 31, 33, 34, 38, 39] further indicating the high quality of our grown crystals.

Surprisingly, with a magnetic field of 7 T applied perpendicular to the *ab* plane, we observed a high in-plane magnetoresistivity up to 22% at 5 K. This is quite unusual, as both theoretical[43,44] and angle-resolved photoemission spectroscopy studies[45] have suggested that α-BaCu$_2$As$_2$ behaves as a simple *sp* metal with weak electronic correlations and Cu$^{1+}$ oxidation states. A Cu$^{1+}$ oxidation state corresponds to a completely filled 3d$^{10}$ configuration, which should not have any net spin moment. A closer look and more careful analysis of the resistivity of our β-BaCu$_2$As$_2$ indicates that a fairly weak resistivity anomaly can be deduced by the first derivation of the temperature-dependent resistivity data at 43 K, as shown in inset of Fig. 4b. This coincides with the magnetoresistivity data where the magnetoresistance starts to emerge below 50 K, as shown in Fig. 4b. Through detailed fitting with a Fermi liquid model in the low-temperature regime using ρ=ρ$_0$+AT$^2$ (Fig. 4c), we obtained ρ$_0$=1.70(1) μΩ cm and A= 0.0009(1) μΩ cm/K$^2$. However, we found the existence of a small deviation from the Fermi liquid fitting at low temperature below 10K, which suggests possible existence of spin fluctuation in this system.[42,57,58] Magnetic susceptibility, shown in Fig. 4d, is nearly flat between 50 K and 300K but increases rapidly when temperature is below 50 K. A weak anomaly is also observed at around 40 K by taking the first derivation of susceptibility data (inset, Fig. 4d), which further support the possible existence of spin fluctuations in the system.

α-BaCu$_2$As$_2$ has already been demonstrated to be nonmagnetic[46], and it is analogous to the nonmagnetic collapsed tetragonal phases in iron pnictide superconductors resulting from strong interlayer interaction. However, in β-BaCu$_2$As$_2$, the incorporation of CaBe$_2$Ge$_2$-type blocks into the structure will influence the interlayer As-As bonding as the interlayer interaction in the CaBe$_2$Ge$_2$-type structure is between Cu and As atoms rather than As-As, which we speculate will drive the Cu state away from the filled 3d$^{10}$ configuration and cause the spin fluctuation in this new phase. This spin fluctuation might become magnified, or new types of magnetic order could emerge, if one could further chemically dope this phase through either nonmagnetic or magnetic ions to tune the interlayer As-As interactions, as seen in the other 122 systems.[59-61] This will be the subject of future studies.

## CONCLUSION

In conclusion, we have reported a completely new polymorphic phase of β-BaCu$_2$As$_2$ with a much larger *c* lattice parameter than that of α-BaCu$_2$As$_2$. This new phase is an ordered intergrowth structure with CaBe$_2$Ge$_2$-type blocks sandwiched between ThCr$_2$Si$_2$-type blocks that retains body-centered symmetry (*I4/mmm*), as confirmed by X-ray single-crystal diffraction, TEM and STEM studies, and comparisons between our intergrowth structure and the simpler ThCr$_2$Si$_2$-type structure of α-BaCu$_2$As$_2$. This new phase displays unusual magnetoresistivity up to 22% at 5 K and under a magnetic field of 7 T, which suggests the existence of spin fluctuation in this system. Our results indicate a route for the discovery of new polymorphic structures through flux and temperature control during material synthesis.

AUTHOR INFORMATION
**Corresponding Authors**
*E-mail: cwchu@uh.edu (P. C. W. Chu)
*E-mail: blv@utdallas.edu (B. Lv)
**ORCID**
Hanlin Wu: 0000-0002-7920-3868
Bing Lv: 0000-0002-9491-5177
**Notes**
The authors declare no competing financial interest

## ACKNOWLEDGMENT

This work at University of Texas at Dallas is supported by US Air Force Office of Scientific Research Grant Nos. FA9550-15-1-0236 and FA9550-19-1-0037. This project is also partially funded by NSF-DMREF-1921581, and the University of Texas at Dallas Office of Research through the Core Facility Voucher and Seed Program for Interdisciplinary Research (SPIRe) Program. The work performed at the Texas Center for Superconductivity at the University of Houston is supported in part by US AFOSR, the T. L. L. Temple Foundation, the John J. and Rebecca Moores Endowment, and the State of Texas through the Texas Center for Superconductivity. MK was supported in part by the Louis Beecherl, Jr. Endowment Funds and Global Research and Development Center Program (2018K1A4A3A01064272) and Brain Pool Program (2019H1D3A2A01061938) through the National Research Foundation of Korea (NRF) funded by the Ministry of Science and ICT.